\documentclass[twocolumn,prl,aps,epsf,raggedbottom,showpacs]{revtex4}
\RequirePackage[usenames]{pstcol}
\usepackage{graphicx}
\usepackage{amsmath}
\usepackage{rotating}
\usepackage{longtable}

\begin{document}

\title{Search for Time-Reversal Violation in the
\boldmath$\beta$\unboldmath-Decay of
Polarized $^8$Li Nuclei}

\author{R.~Huber, J.~Lang, S.~Navert\footnote{Permanent address:
Swiss Nuclear Safety Inspectorate,
CH-5232 Villigen HSK, Switzerland}, J.~Sromicki}
\affiliation{Institut f\"ur Teilchenphysik, Eidgen\"ossische
Technische Hochschule, CH-8093 Z\"urich, Switzerland}
\author{K.~Bodek, St.~Kistryn, J.~Zejma}
\affiliation{Instytut Fizyki, Uniwersytet Jagiello\'nski,
PL-30059 Krak\'ow, Poland}
\author{O.~Naviliat-Cuncic}
\affiliation{Laboratoire de Physique Corpusculaire de Caen,
ISMRA-IN2P3,
14050 Caen, France}
\author{E.~Stephan}
\affiliation{Instytut Fizyki, Uniwersytet \'Sl\c{a}ski, PL-40007
Katowice, Poland}
\author{W.~Haeberli}
\affiliation{University of Wisconsin, Madison, Wisconsin 53706, USA}

\date{\today}
\begin{abstract} The transverse polarization of  electrons
emitted in the $\beta$-decay of polarized $^8$Li nuclei has been
measured. For the time reversal violating triple correlation
parameter we find $R=(0.9\pm 2.2)\times 10^{-3}$. This result is
in agreement with the standard model and yields improved
constraints on exotic tensor contributions to the weak interaction.
It also provides  a new limit on the
mass of a possible scalar leptoquark,
$m_{LQ} > 560 \;\rm{GeV}/c^2$ (90\% C.L.).
\end{abstract}


\maketitle

Following the surprising discovery of the small violation of
the $CP$ symmetry in the neutral kaon system almost 40 years
ago~\cite{Christenson}, larger signatures
have been observed more recently in the decay of neutral $B$
mesons~\cite{AubertAbe}.
According to the {\em CPT} theorem any violation of
the combined particle-antiparticle and space inversion
symmetries $(CP)$ is equivalent to a violation of time
reversal symmetry {\em (T)}. However, only  recently a
measurement of a difference between the rate of a process and its
inverse and thus the first {\em direct} confirmation of a time
reversal violation has been published \cite{Ang}.
All the observations performed so far  could be accommodated
within the standard model (SM) through Cabibbo - Kobayashi -
Maskawa mixing of the quark states. It is not yet clear
whether the small {\em direct CP} violation measured in the decay
of the $K^0$ (the $\epsilon'/\epsilon$-parameter) \cite{Bat} can also be
explained by the same mechanism, since hadronic corrections cannot
be calculated with sufficient precision.
Furthermore, the amplitude of $CP$ violation
due to mixing of the quark states is by several orders of magnitude
too small to explain the  matter-antimatter asymmetry of the Universe.
Despite huge efforts during the past three decades, no
{\em T} or {\em CP} violation has ever been observed outside the
$K^0$ or $B^0$ systems. If the only source of {\em CP}-violation
would be the one offered by the standard model, effects in $\beta$-decay
would be second order in the weak interaction and therefore
vanishingly
small. Thus any observation of a time reversal violation in such a
process would be the first unambiguous signal of {\em new physics}
beyond the SM.

It has been shown in Refs.~\cite{Deu,Her} that
the accurate determination of the transverse
polarization of the electrons emitted in the decay of polarized
nuclei can provide such a {\em precision test}.
An appropriate experiment could yield a
new limit for a possible time reversal {\em (T)} non-invariant
tensor-coupling. In renormalizable gauge-theories the only 
mechanism that could generate such a coupling at the tree level
is the exchange of a spin-0-leptoquark \cite{Her}.

We present here the final result of a measurement of the
transverse polarization of electrons emitted in the decay of
polarized $^{8}$Li nuclei. A detailed
description of the measuring principle and of the experimental setup
have been reported previously \cite{Sro}.
We describe here the modifications of the experimental setup
which allowed to obtain a considerably higher precision. 
The most important improvements are a doubling of the
$^{8}$Li polarization and an order of magnitude reduction in
background events.

The terms in the allowed $\beta$-decay rate function which are
relevant for the following discussion are the ones given by
the usual parity violating $\beta$ asymmetry parameter $A$ and
the parity and time reversal violating triple correlation
coefficient $R$ \cite{Jac}.
\begin{equation}\label{eq:rate}
   W \propto \left( 1 + A\;
   \frac{\vec{J}\cdot\vec{p_e}}{E_e}
   + R\;\frac{\vec{J}\cdot
   (\vec{p_e}\times\vec{\sigma_e})}{E_e}
   + ... \right).
\end{equation}
In this expression $\vec{J}$ denotes the initial nuclear polarization
whereas $\vec{p}_e$, $E_e$ and $\vec{\sigma}_e$ correspond to the
momentum, energy and spin of the electron, respectively.

\begin{figure}[htb]
  \begin{center}
  \includegraphics[height=6cm]{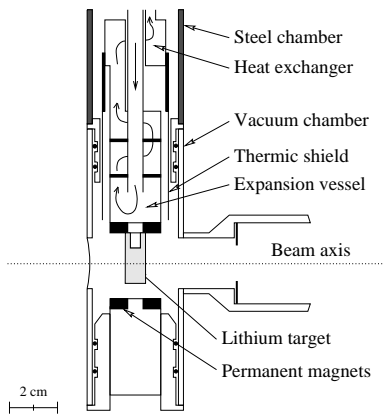}
  \end{center}
  \caption{\label{FIG_1} Simplified  view of the $^7$Li target
cryostat.
The horizontal access of the beam is indicated on the right.
The outer diameter of the cryostat is 40~mm}
  \label{fig:target}
\end{figure}

The experiment was performed at the Paul Scherrer Institute, 
Villigen, Switzerland.
Polarized $^8$Li nuclei were produced in the reaction
  $^7\textrm{Li}(\vec{\textrm{d}},\textrm{p})^8\vec{\textrm{Li}}$
using a 0.9 $\mu$A beam  of 10 MeV vector-polarized deuterons from a
polarized-ion source based on the atomic-beam method. Compared to the earlier
experiments, the beam polarization was improved by use of radio-frequency
transitions on the atomic beam which, ideally, yield reversible vector
polarizations of $\pm 1$~\cite{Rob}. The transitions were
adjusted by optimizing the asymmetry in  the
  $^{12}\textrm{C}(\vec{\textrm{d}},\textrm{p})^{13}\textrm{C}$
  reaction.  The target, a 5 mm diameter rod
of 99.9 \% enriched $^7\textrm{Li}$ metal, was cooled with liquid
helium and placed in a magnetic field of $B_y$ = 7 mT
(c.f. fig. \ref{FIG_1}) to achieve a long polarization relaxation
time $(T \ge 20\, \textrm{s}$),
an order of magnitude longer than the mean decay time $(\tau = 1.21
\,\textrm{s} )$.

\begin{figure}[htb]
  \begin{center}
  \includegraphics[height=4cm]{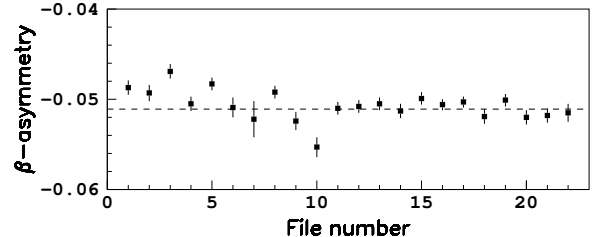}
  \end{center}
  \caption{\label{FIG_2} Beta asymmetry measured with the
  $^8\textrm{Li}$ polarization monitor as a function of file (run)
number}
\label{fig:asy}
\end{figure}

The polarization of the $^8$Li target was deduced from the beta decay asymmetry
detected by  plastic scintillator telescopes placed at
$\theta = 45^\circ$ and $135^\circ$ with respect to the 
$^8\textrm{Li}$ polarization axis.The Li target was viewed by the detectors
through small holes in the shielding.
The evolution of the asymmetry during the experiment
as measured by one of the telescopes is shown in fig. \ref{fig:asy}.
 From the super-ratios of the counting rates of the two telescopes
and the value of the asymmetry parameter for a $J_i = J_f$ pure
Gamow-Teller transition ($A = -1/3$) the value of the
$^8\textrm{Li}$ polarization, averaged over the whole measurement
period, was found to be $|\vec{J}| = 0.21\pm 0.01$, which represents
a factor 2 improvement over the earlier experiment \cite{Sro}.

\begin{figure}[htb]
  \begin{center}
\includegraphics[height=6cm]{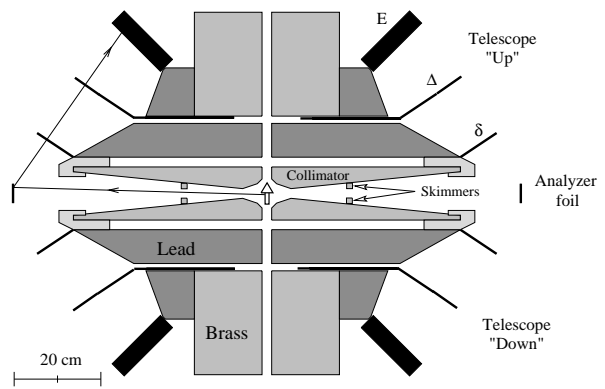}
  \end{center}
  \caption{Vertical cross section through the Mott
  polarimeter. The direction of the incident polarized deuteron beam is
  perpendicular to the figure. The central arrow indicates the
direction  of the $^8$Li spin in the target. A trajectory of an electron
  scattered on the lead foil is also shown.
  The diameter of the analyzer foil ring is 1.1~m.}
  \label{fig:polarimeter}
\end{figure}

The transverse polarization of the electrons emitted in the decay
$^8\textrm{Li}\rightarrow {^8\textrm{Be}} (2.9\,\textrm{MeV}) + e^- +
\bar{\nu}_e$  was deduced from the measured
asymmetry in Mott scattering at backward angles. A  lead foil
of 105 mg/cm$^2$  thickness was used as analyzer.
To obtain a large solid angle the asymmetry detectors
were arranged in a cylindrical geometry around the $^8$Li
polarization axis (c.f. fig.~\ref{fig:polarimeter}).
In practice, the detector was made of four separate azimuthal
segments, each containing an upper and a lower telescope, which
provided four independent measurements of the electron polarization.
The telescopes  consisted of two  thin transmission
scintillators ($\delta$, $\Delta$) followed
by a stopping scintillator ($E$).
Much attention was paid to the passive shielding of the
detectors against background radiation produced
in the target neighborhood.  In the central part
of the apparatus, additional cylindrical electron skimmers
were mounted on the  brass collimators.  They reduced
significantly multiple scattering of electrons on the collimator
plates. All other shieldings with large atomic numbers have been 
covered with plastic to reduce the scattering of electrons.

The modifications of the apparatus mentioned above led to a
signal-to-background ratio between 75 and 150 in the different
segments of the polarimeter. This is an improvement of an order of
magnitude compared with the conditions in the earlier stage of the
experiment \cite{sro, Sro}.

The measurement was performed in a cyclic fashion with a 0.33 s
activation of the target followed by a measuring period divided
into 32 intervals of 33 ms each.  The sign of the beam
polarization $p_y$ was reversed every four activation/measurement
cycles.  The data were accumulated separately for eight telescope
sectors in form of two-dimensional spectra (time vs. energy) and
stored for off-line analysis.  24 runs of 2 hours duration with
the scattering foil in place ({\it foil-in}) and 2 runs with foil
removed ({\it foil-out}) were taken.  The {\it foil-out} runs
provided the information on the intensity and asymmetry of the
background that affected each telescope.

It has been shown previously \cite{Sro} that different methods
of analyzing the data lead to virtually the same results.
The data collected in the current
experiment were analyzed twice. In the first
procedure, the polarization of the $^8$Li source was assumed to be
constant in time.  In the second one, each file of raw data,
corresponding to a single run, was associated with the
polarization value measured for this particular run.
Again, the final results and their statistical consistency
are almost identical which indicates that the asymmetry variations
shown in Fig.~\ref{fig:asy} have no significant effect.
In both analyses  the  correlation coefficient R was
calculated file-by-file, separately
for four pairs of telescopes and then the weighted average of the four
telescopes was calculated.

Table \ref{tab:results} summarizes the evolution of the
improvements achieved during this project. 
The table includes in columns I-IV
the results obtained previously \cite{sro, Sro}. Series V and
VI were obtained with the modified setup
presented here, except that for series V only some of the improvements had
been implemented.

The most important steps in the treatment of the data and the
corrections were the following:

{\it Analyzing power.} The spin dependence (analyzing power
or Sherman function) in the scattering of 14 MeV electrons from
Pb as a function of scattering angle and foil thickness
was investigated in a separate experiment at the Mainz Microtron, c.f.
Ref. \cite{Dep}.
In order to minimize background and multiple scattering
effects in the lead foil, the lower threshold for the energy of
scattered electrons was set to 4 MeV. (The $\beta$-endpoint energy
is $13.1$ MeV).
The thickness of the lead foil
was chosen to optimize the count rate while limiting the decrease in
effective analyzing power by multiple scattering in the foil.  In the
present experiment the mean value of the analyzing power was ${\cal
A} = 0.08\pm 0.01$.

{\it Cuts in time.} The first time channel (33 ms) after each
activation period was disregarded since the rate recorded by all
detectors in this time interval was higher by an order of magnitude
than that in the next channel.  This excess was attributed
to short lived background activities after irradiation.

{\it Gain corrections and energy cuts.} The amplification of the
detectors was monitored with LED flashes fired at the
beginning of each time channel.  The amplifications were
always observed to be about 5 \% larger at the beginning of
the counting interval.  This effect is very similar for the two
polarization states of the $^8$Li source.  As a consequence, the gain
variation contribution to the uncertainty of the R-coefficient is
strongly suppressed.

{\it Decay background.} The fraction of events not originating
from electrons scattered by the analyzer foil was $0.008 \pm
0.002$.  It was deduced from the  measurements with
removed scattering foil.  The corresponding correction was
calculated for each pair of telescopes separately.

{\it Accidentals, dead time and pile up.} The rate of accidental
coincidences was determined by inserting  delay lines into
the input channels of the master coincidence unit.  The rate of accidentals
compared to the regular {\it foil-in} rate was only $\approx
0.0009$.  The correction was applied separately for each pair of telescopes.
The effects from dead time and pile up are very small and are combined in
Tab.~\ref{TAB_1} with accidentals.

\begin{widetext}
\begin{center}
\begin{table*}
  \caption{\label{TAB_1} Deduced values of the R-coefficient including
  applied corrections in units of $10^{-3}$ for the present and five
  former results.}
  \vspace{2mm}
  \begin{tabular*}{150mm}[b]{@{\extracolsep\fill}|p{3cm}|r|r|r|r|r|r|}

  \hline\hline
  ~Series & I Ref.\cite{sro} & II Ref.\cite{sro} &
III Ref.\cite{Sro} & IV Ref.\cite{Sro}
& V Ref.\cite{Nav} & VI present \\
  \hline\hline
  ~Raw data & $12 \pm 34$ & $9 \pm 12$ & $-13 \pm 7$ &
     $-8 \pm 4$ & $-3.6 \pm 6.1$ & $0.7 \pm 2.6$ \\
  ~Decay background & $-39 \pm 26$ & $-13 \pm 8$ & $6 \pm 4$ &
     $0 \pm 2$ & $2.5 \pm 1.1$ & $-0.5 \pm 0.9$ \\
  ~Gain shifts & $-1 \pm 3$ & $-1 \pm 2$ & $-1 \pm 1$ &
     $-1 \pm 1$ & $-0.7 \pm 1.6$ & $0.0 \pm 0.4$ \\
  ~Accidentals & $-7 \pm 8$ & $-1 \pm 2$ & $2 \pm 2$ &
     $1 \pm 1$ & $-1.6 \pm 1.1$ & $-0.4 \pm 0.5$ \\
  ~Beta asymmetry & $11 \pm 5$ & $13 \pm 2$ & $10 \pm 1$ &
     $6 \pm 1$ & $5.5 \pm 0.5$ & $2.5 \pm 0.7$ \\
  \hline
  ~Result & $-24 \pm 44$ & $7 \pm 15$ & $4 \pm 8$ &
     $-2 \pm 5$ & $2.1 \pm 6.5$ & $2.3 \pm 2.9$ \\
  \hline\hline
  \end{tabular*}
\label{tab:results}
\end{table*}
\end{center}
\end{widetext}

{\it Beta asymmetry.} This nonstatistical effect is associated
with the nonuniform illumination of the scattering foil.  It
arises from an interplay between the asymmetric angular
distribution of decay electrons from polarized nuclei and the finite
geometry of the apparatus, in particular, the scattering foil.
(see Ref. \cite{Sro}).
This is a small
but measurable effect which does not cancel under inversion
of the nuclear spin $\vec{J}$ (see Eq.~(\ref{eq:rate})).
  Since the magnitude of this effect is proportional to the
square of the height $h^2$ of the scattering foil, it was reduced in
series V and VI by using a lead foil with a smaller height
(40 mm and 25 mm, instead of 50 - 70 mm as in the previous
experiments).
The associated correction was determined by replacing the
regular scattering foil used during the measurement by one
of smaller size which was moved over the region covered by
the regular foil \cite{Nav}.

The weighted average value of the R-correlation deduced from all
the six runs including the last two with the modified setup is
\begin{equation}
\label{eq:Rres}
   R_M = (1.6 \pm 2.2) \times 10^{-3}
\end{equation}

{\it Final state interaction.} The effects of the final state
interaction (FSI), which can mimic the genuine time reversal
violation in the $R$ correlation, are exceptionally small for the
$^8$Li decay.  An average value of the FSI correction over the
energy range of detected electrons and calculated in the point
nucleus approximation \cite{Jac}, amounts to $R_{\mathrm FSI}=0.7
\times 10^{-3}$ and is known with 10 \% accuracy.  This accuracy is
adequate for our experiment although more precise calculations of
the FSI correction are possible \cite{Fsi2,Fsi3}.

Taking into account the FSI effects we obtain the
time reversal violating part $R$ from the measured correlation
$R_{M}$:
\begin{equation}
   R = R_M - R_{\mathrm FSI} = (0.9 \pm 2.2) \times 10^{-3}.
\end{equation}
This result is consistent with time reversal invariance.

One of the distinct advantages of the decay  $^8\textrm{Li}
\rightarrow {^8\textrm{Be}} (2.9\,\textrm{MeV}) + e^- +
\bar{\nu}_e$ is its simple spin-isospin structure ($\Delta J = 0,
\Delta T = 1$). The Fermi matrix element vanishes and one has an
essentially pure, allowed Gamow-Teller transition. In this case
the $R$-parameter depends only on the tensor interaction:
\begin{equation}
R = -\frac{4}{3}\Im{\mathit m}{\left(\frac{g_T}{g_A}a^T_{RL}\right)} =
\frac{1}{3}\Im{\mathit m}\left(\frac{C_T + C_T'}{C_A}\right)
\end{equation}
We use the helicity projection constants  ($g_T, g_A,
a^T_{RL}$) as defined in Ref.~\cite{Deu}. They are analogous to the
ones used in the Review of Particle Properties for pure leptonic
transitions \cite{Fet}. $C_T, C_T'$ are the unknown tensorial coupling
constants and $|C_A| \approx 1.26$  is the
axial-vector coupling constant as defined in
Ref.~\cite{Jac}. If we use the same normalization as
Ref.~\cite{Deu} for the tensorial coupling constant, $g_T/g_A = 1$,
we get the new limit
\begin{equation}
   \Im{\mathit m} (a^T_{RL}) = -0.0007 \pm 0.0016 \qquad (1 \sigma)
\end{equation}
\begin{equation}
  |\Im{\mathit m}(a^T_{RL})| < 0.0029 \qquad (90\%\;\rm{ C. L.})
\end{equation}
which has roughly the same precision as the best determination of
the real part, $|\Re{\mathit e} (a^T_{RL})| < 0.0035 $ (90\%\;
C.L.)\cite{Deu}. We therefore also obtain an improved limit for
\begin{equation}
|a_{RL}^T| < 0.0044 \qquad (90\%\;\rm {C.L.})
\end{equation}
Finally, for a possible coupling with scalar leptoquarks, we
obtain using a relation from Ref.~\cite{Her}:
\begin{equation}
\frac{m_{LQ}}{|h_L h_R^*|^{\frac{1}{2}}} = \left|\frac{\sqrt{2}}{8 G_F
a^T_{LR}}\right|^{\frac{1}{2}} > 1.8 \rm{GeV \quad at\quad 90\% \;\rm{
C.L}.}
\end{equation}
$h_R$ and $h_L$ are the (unknown) coupling constants of the
leptoquarks and $G_F$ the Fermi coupling constant. If we assume
``canonical'' values for $h_{L,R} = \sqrt{4 \pi \alpha_{elm}}
\approx 0.3$
  we get an estimate for the mass of a possible scalar
leptoquark with charge $Q=\pm 2/3 e$ of $m_{LQ} > 560$
GeV$/c^2$. This has to be compared to limits from high energy
experiments, which yield in general
lower limits of approx. 300 GeV/$c^2$ \cite{Lep}.

\end{document}